\documentstyle[aps,prl,multicol,epsf]{revtex}
\begin{document}
\draft
\title{Bilayer Quantum Hall States at $\nu=1$ and
Coulomb Drag}

\author{Yong Baek Kim$^1$, Chetan Nayak$^2$, Eugene Demler$^3$,
N. Read$^4$, and S. Das Sarma$^5$}
\address{$^1$ Department of Physics, The Ohio State University,
Columbus, OH 43210\\ $^2$ Department of Physics, University of
California, Los Angeles, CA 90095\\ $^3$ Department of Physics,
Harvard University, Cambridge, MA 02138\\ $^4$ Department of
Physics, Yale University, P.O. Box 208120, New Haven, CT 06520\\
$^5$ Department of Physics, University of Maryland, College Park,
MD 20742 }

\date{\today}
\maketitle

\begin{abstract}
We consider a number of strongly-correlated quantum Hall states
which are likely to be realized in bilayer quantum Hall systems at
total Landau level filling fraction ${\nu_T}=1$. One state, the
$(3,3,-1)$ state, can occur as an instability of a compressible
state in the large $d/l_B$ limit, where $d$ and $l_B$ are the
interlayer distance and magnetic length, respectively. This state
has a hierarchical descendent which is interlayer coherent.
Another interlayer coherent state, which is expected in the small
$d/l_B$ limit is the well-known Halperin $(1,1,1)$ state. Using
the concept of composite fermion pairing, we discuss the
wavefunctions which describe these states. We construct a phase
diagram using the Chern-Simons Landau-Ginzburg theory and discuss
the transitions between the various phases. We propose that the
longitudinal and Hall drag resistivities can be used together with
interlayer tunneling to experimentally distinguish these
different quantum Hall states. Our work indicates the bilayer
${\nu_T}=1$ quantum Hall phase diagram to be considerably richer
than that assumed so far in the literature.
\end{abstract}


\begin{multicols}{2}

\section{Introduction}

Bilayer quantum Hall systems at total Landau level filling factor
${\nu_T}=1$, i.e.\ $\nu=1/2$ in each layer \cite{footnote_1},
allow for novel interlayer coherent phases. These phases have
attracted a great deal of theoretical and experimental attention
\cite{Eisenstein97} over the last sixteen years, dating back to a
seminal paper by Halperin \cite{Halperin}, in which the
multicomponent generalization of the Laughlin wavefunction was
first considered in a rather general context. In particular, there
is strong experimental evidence \cite{Eisenstein97,eisenstein} and
compelling theoretical basis \cite{Eisenstein97,Wen92} to believe
that a spin-polarized bilayer $\nu={1\over 2}$ quantum Hall system
would have a novel spontaneous interlayer coherent incompressible
phase for small values of the interlayer separation $d$, even in
the absence of interlayer tunneling. (We consider only the
situation without any interlayer tunneling in this paper -- our
considerations also apply to the physical situation with weak
interlayer tunneling. The situation with strong interlayer
tunneling is trivial by virtue of the tunneling-induced
symmetric-antisymmetric single-particle tunneling gap which leads
to the usual ${\nu_T}=1$ quantum Hall state in the symmetric
band.) In the limit $d\rightarrow\infty$, however, one expects two
decoupled layers, each with $\nu={1\over 2}$, and hence no
quantized Hall state. The phase diagram for this compressible to
incompressible quantum phase transition in $\nu={1\over 2}$
bilayer systems has been studied extensively in the literature
\cite{Eisenstein97}, but we still do not have a complete
qualitative understanding of the detailed nature of this
transition. In particular, one does not know how different kinds
of incompressible (and compressible) phases compete as system
parameters (e.g. $d$) are tuned, and how to distinguish among
possible competing incompressible phases. In this paper, we
revisit this issue by arguing that, in principle, there are
several interesting and nontrivial quantum Hall phases in the
${\nu_T}=1$ bilayer system which could be systematically probed
via interlayer drag experiments carried out at various values of
the interlayer separation $d$. One goal of our paper is to
describe and discuss the rich ${\nu_T}=1$ bilayer quantum Hall
phase diagram using the Chern-Simons Landau-Ginzburg approach
\cite{Zhang89,Read89,Read90,Blok90,Moon95}.

Among the more interesting quantum Hall phases are the so-called
paired Hall states, which have been extensively
studied theoretically \cite{Halperin,Moore91,bonesteel}.
In these states, the composite fermions form a
superconducting paired state in which two
composite fermions bind into effective Cooper pairs
which condense into a ground state
analogous to the BCS state. The well-studied
Moore-Read Pfaffian state \cite{Moore91} is a
spin-polarized version of such a paired Hall
state for a single layer system.
Bilayer paired Hall states have been discussed
earlier in the literature in the context of
$\nu={1\over 2}$ (and $\nu={1\over 4}$ systems),
but no definitive idea has emerged regarding
their experimental observability or
their relation to the more intensively studied
(and robust) $(1,1,1)$ state
\cite{Eisenstein97,Halperin,Wen92}.
The main purpose of the current paper is to critically discuss
the possible existence of paired $\nu={1\over 2}$
bilayer Hall states which,
we argue, are distinguishable from
the better-studied $(1,1,1)$ incompressible
state (as well as from compressible states)
through interlayer drag experiments. The transitions
between these states are described by a
variety of Landau-Ginzburg theories.
Given the great current interest \cite{eisenstein,demler}
in the physics of $\nu={1\over 2}$ bilayer systems,
we believe that the results presented in this paper
could shed considerable light on the nature of the possible
quantum phase transitions in bilayer systems.

In sections II, III, and IV of this paper, we consider the
possible bilayer $\nu={1\over 2}$ quantum Hall phases in the
parameter regimes $d\gg {\ell_B}$ (section II), $d> {\ell_B}$
(section III), and $d \stackrel{<}{\sim} {\ell_B}$ (section IV),
where ${\ell_B}$ is the magnetic length, which sets the scale for
intralayer correlations. We argue that the likely ground states in
these three regimes are, respectively, compressible
(Fermi-liquid-like) states ($d\gg {\ell_B}$), paired Hall states
($d> {\ell_B}$), and $(1,1,1)$ states ($d\sim {\ell_B}$). In
section V, we discuss the transition between the $d> {\ell_B}$ and
$d \stackrel{<}{\sim} {\ell_B}$ limits, and introduce yet another
state, a hierarchical descendent of the (3,3,-1) state with
interlayer coherence. We conclude in section VI with a critical
discussion of the various drag resistivities which we argue can,
in principle, distinguish among these phases and could be used to
study bilayer quantum phase transitions experimentally.

\section{$\lowercase{d}\gg \ell_B$: Compressible State}

Let us consider a bilayer system in which each layer has
filling factor $\nu=1/2$ in the limit that the layer separation
$d$ is much larger than the typical interparticle spacing,
which is of the order of the magnetic length $\ell_B$.
As a starting point, we will model the system
by two almost independent Fermi-liquid-like
compressible states, one in each layer.
There are two alternative and complementary
descriptions of compressible states at even-denominator
filling fractions. We will briefly recapitulate
some features of both, as they will inform
the following discussion of paired states.

One description of a Fermi-liquid-like
compressible state at $\nu=1/2$ is based on the
lowest Landau level wavefunction \cite{rezayi}:
\begin{equation}
\label{eqn:RR_wavefcn}
\Psi_{1/2} (\{z_i\})
= {\cal P}_{LLL}\: {\rm Det} M
\prod_{i<j} (z_i  - z_j )^2 \ ,
\end{equation}
where $M$ has the matrix elements
$M_{ij} = e^{i {\bf k}_i \cdot
{\bf r}_j}$. Here, ${\bf r}_i$
is the position of electron $i$. The
${\bf k}_i$s are parameters which
are chosen so that
the total energy of the system is minimized.
We will discuss their interpretation below.
${\cal P}_{LLL}$ projects states into the lowest Landau level;
it has the following action \cite{rezayi,girvin}
\begin{equation}
{\cal P}_{LLL} e^{i {\bf k}_i \cdot
{\bf r}_j} {\cal P}_{LLL} =
e^{i {\bf k}_i \cdot
{\bf R}_j} \ ,
\end{equation}
where ${\bf R}_i$ are the guiding center coordinates of the
electrons.

The corresponding wavefunction of the double-layer system
at ${\nu_T}=1$ can be written as
\begin{equation}
\Psi_{\rm compresible}
= \Psi_{1/2} (\{z^{\uparrow}_i\})
\Psi_{1/2} (\{z^{\downarrow}_i\}) \ ,
\end{equation}
where $\alpha = \uparrow, \downarrow$ label the
two layers.

The lowest Landau level constraint and Fermi statistics
displace the electrons from the
correlation holes -- i.e. zeroes of
the wavefunction or, equivalently, vortices
represented by the $\prod_{i<j} (z_i-z_j)^2$
factor in the wavefunction.
The ${\rm Det} M$
factor is necessary to ensure Fermi statistics;
it is a displacement operator \cite{read94} because
${\bf k}_i \cdot {\bf r}_j$ acts as
${\bar k}_i z_j + k_i {\partial \over \partial z_j}$ in the
lowest Landau level.
Thus the composite fermion made of an electron and
two correlation
holes has a dipolar structure.
Using these ideas, the system
can be described as a collection of dipolar `composite
fermions' in which each dipolar fermion consists of an electron
and the corresponding correlation holes
\cite{read94,shankar,stern,read98}.
To each composite fermion we assign a
${\bf k}_i$ which is equal in magnitude
(in units of $\ell_B$)
and perpendicular to the displacement
between the electron and its correlation holes.
By Fermi statistics, the ${\bf k}_i$s
must be distinct. The structure of a dipolar composite fermion
is given schematically in Fig.1. The energy of a composite
fermion increases with ${\bf k}_i$,
so the ground state is a filled Fermi
sea in ${\bf k}$-space.
In the long wavelength limit,
the total energy of these dipolar composite fermions can be
approximately written as $\sum_i k^2_i / 2m^*$ with
effective mass $m^*$ which is
determined by the interaction
potential \cite{shankar,read98}.

An alternative formulation arises from the observation that an
electron may be represented by a `composite fermion' together with
a Chern-Simons gauge field which attaches two fictitious flux
quanta to each fermion \cite{jain,CS,HLR}. This representation is
mathematically equivalent to the original one, but it naturally
suggests another approximation. The composite fermions see zero
average magnetic field due to the cancellation between the
external magnetic field and the average fictitious magnetic field
coming from the fictitious flux quanta. Consequently, the system
can be described as an almost free (composite) fermion system in
zero effective magnetic field. In this approach, the fictitious
flux quanta are introduced to represent the phase winding of the
electron wavefunction around correlation holes associated with the
positions of other electrons. The ${\rm Det} [e^{i {\bf k}_i \cdot
{\bf R}_j}]$ factor in (\ref{eqn:RR_wavefcn}) can be interpreted
as the wavefunction of the almost free composite fermion system.
\cite{rezayi,HLR} Then ${\bf k}_i$s can be regarded as the
``kinetic momenta'' of the composite fermions. In the
long-wavelength, low-frequency limit, the two formulations are
equivalent.

\begin{figure}
\vspace{-0.5truecm}
\center
\centerline{\epsfysize=1.2in
\epsfbox{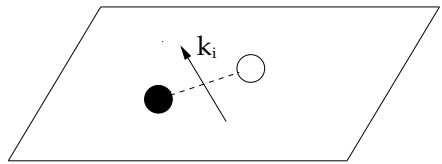}}
\vspace{-1.0truecm}
\begin{minipage}[t]{8.1cm}
\caption{Schematic picture of the dipolar composite
fermion. The black and white dots represent the
electron and vortex respectively. The ``wavevector''
${\bf k}_i$ is perpendicular to the dipole moment.
}
\label{fig1}
\end{minipage}
\end{figure}

Note that in addition to the Fermi-liquid-like $\nu=1/2$
compressible state in each layer one could have, in principle,
other compressible states (e.g.\ CDW or Wigner crystal) in the
$d\gg {\ell_B}$ limit depending on the details of interaction and
Landau-level coupling. In addition, strong disorder will lead to
localized insulating states. We do not consider these
possibilities in this paper.

\section{$\lowercase{d}>\ell_B$: Pairing}

We will now implement the latter formulation
in a double-layer system, and show
that the compressible state
has a strong pairing instability \cite{bonesteel}.
We introduce two composite fermion
fields $\psi_\alpha$ and two Chern-Simons
gauge fields ${\bf a}_\alpha$, $\alpha=\uparrow,\downarrow$.
The Hamiltonian is
\begin{eqnarray}
H &=& H_0 + H_I  \ , \cr
H_0 &=& \int d^2 r \sum_{\alpha = \uparrow, \downarrow}
{1 \over 2m^*} \psi^{\dagger}_{\alpha}
(\nabla - {\bf a}_{\alpha})^2 \psi_{\alpha} \ , \cr
H_I &=& \int d^2 r \int d^2 r' \
\delta \rho_{\alpha} ({\bf r})
V_{\alpha \beta} ({\bf r}-{\bf r'})
\delta \rho_{\beta} ({\bf r'})
\end{eqnarray}
with the constraints $\nabla \times {\bf a}_{\alpha}
= 2 \pi {\tilde \phi} \delta \rho_{\alpha} ({\bf r})$.
${\tilde \phi} = 2$ if the filling factor of each layer is $\nu=1/2$.
Here $\delta \rho ({\bf r}) = \rho ({\bf r})-{\bar \rho}$
is the density disturbance measured
from the average value ${\bar \rho}$.
The interaction potential is given by
$V_{\uparrow \uparrow} = V_{\downarrow \downarrow} = e^2/\varepsilon r$
and $V_{\uparrow \downarrow} = V_{\downarrow \uparrow} =
e^2 / \varepsilon \sqrt{r^2 + d^2}$
where $\varepsilon$ is the dielectric
constant.

It is convenient to change the gauge field variables to
${\bf a}_{\pm} = {1 \over 2}
({\bf a}_{\uparrow} \pm {\bf a}_{\downarrow})$.
Then $H_0$ can be rewritten as
\begin{eqnarray}
\label{eqn:H_0}
H_0 &=& \int d^2 r \left [
{1 \over 2m^*} \psi^{\dagger}_{\uparrow}
(\nabla - {\bf a}_+ - {\bf a}_-)^2 \psi_{\uparrow} \right . \cr
&&+ \left . {1 \over 2m^*} \psi^{\dagger}_{\downarrow}
(\nabla - {\bf a}_+ + {\bf a}_-)^2 \psi_{\downarrow}
\right ]
\end{eqnarray}
with $\nabla \times {\bf a}_{\pm} = \pi {\tilde \phi} [\delta
\rho_{\uparrow}({\bf r}) \pm \delta \rho_{\downarrow}({\bf r})]$.
{}From (\ref{eqn:H_0}), we see that $\psi_{\uparrow}$ and
$\psi_{\downarrow}$ have the same gauge ``charges'' for ${\bf
a}_{+}$, but opposite gauge ``charges'' for ${\bf a}_{-}$. As a
result, there will be an attractive interaction between the
composite fermions in different layers via ${\bf a}_-$ and a
repulsive interaction via ${\bf a}_+$. Composite fermions in the
same layer have repulsive interactions. As a result of Coulomb
intractions, the attractive interaction mediated by the ${\bf
a}_-$ gauge field dominates in the low energy limit and there
exists a pairing instability between the composite fermions in
different layers. This result can be understood in physical terms
as follows. The ${\bf a}_-$ and ${\bf a}_+$ fields represent
antisymmetric and symmetric density fluctuations. In the presence
of Coulomb interactions, symmetric density fluctuations are highly
suppressed, but antisymmetric density fluctuations can still be
large. As a result, the dynamic density fluctuations in the
antisymmetric channel becomes more important in the low energy
limit and lead to a pairing instability.

This pairing instability has a natural explanation in the dipolar
composite fermion picture \cite{read94,shankar,stern,read98}. Let
us take a dipolar composite fermion in layer $\uparrow$ with
wavevector ${\bf k}^{\uparrow} = {\bf k}_F$ and a dipolar
composite fermion in layer $\downarrow$ with wavevector ${\bf
k}^{\downarrow}=-{\bf k}_F$. As seen in Fig.\ 2, this
configuration can lower the interlayer Coulomb energy because the
electron in layer $\uparrow$ and the vortex in layer $\downarrow$
can sit on top of each other and vice versa.

\begin{figure}
\center
\centerline{\epsfysize=1.4in
\epsfbox{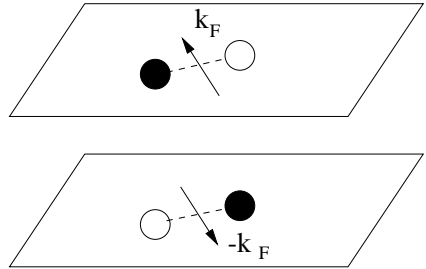}}
\begin{minipage}[t]{8.1cm}
\caption{Pairing of the
``intralayer'' composite fermions.
}
\label{fig2}
\end{minipage}
\end{figure}

This analysis predicts that, at least in principle,
the Fermi-liquid-like compressible state is
always unstable to pairing. In practice,
the pairing gap will be small in the limit
$d\gg \ell_B$ and easily destroyed by disorder.
As a result, we expect the pairing instability
discussed above to be relevant for $d/\ell_B$
not too small. When $d/\ell_B$ become small, on the
other hand, the starting point of two decoupled compressible states
is no longer sensible, and we take a different
starting point, as described in later sections.

The wavefunction of the paired quantum Hall state
constructed in this way can be written as
\begin{eqnarray}
\Psi_{\rm pair} =
\Psi^{\rm cf}_{\rm pair}
\prod_{i>j} (z^{\uparrow}_i-z^{\uparrow}_j)^2
\prod_{k>l} (z^{\downarrow}_k-z^{\downarrow}_l)^2 \ ,
\end{eqnarray}
where
\begin{eqnarray}
\Psi^{\rm cf}_{\rm pair} &=& {\rm Pf} [f(z_i, z_j ; \uparrow,
\downarrow)] \cr &\equiv& {\cal A} [ f(z_1, z_2 ; \uparrow,
\downarrow) f(z_3, z_4; \uparrow, \downarrow) \cdots ] \ ,
\end{eqnarray}
and $f(z_1, z_2 ; \uparrow, \downarrow)$ is the pair wavefunction
which depends on the symmetry of the pairing order parameter.
${\rm Pf} [\cdots]$ denotes the Pfaffian, which is defined in the
second line, with ${\cal A} [\cdots]$ denoting the antisymmetrized
product. Notice that $\Psi_{\rm pair}$ can be regarded as the
product of the wavefunction of the paired composite fermions and
that of the (2,2,0) bosonic Laughlin quantum Hall state.

It is not immediately clear what choice of $f(z_1, z_2 ; \uparrow,
\downarrow)$ is most favorable energetically. In the Chern-Simons
theory of \cite{bonesteel}, there is a pairing instability in all
angular momentum channels \cite{bonesteel}. In a modified
Chern-Simons theory, it was claimed that the leading instability
occurs in the $p$-wave channel \cite{morinari}. These approximate
calculations do not necessarily capture the detailed energetics
which determines the pairing symmetry. Hence, we will not enter
into a discussion of energetics, but limit ourselves to a
discussion of the simplest (and, therefore, likeliest)
possibilities.

The simplest possibility is $p_x - i p_y$ pairing,
\begin{equation}
\Psi_{\rm pair} = {\rm Pf} \left [ {\uparrow_i \downarrow_j +
\downarrow_i \uparrow_j \over z_i - z_j } \right ]
\prod_{i>j} (z^{\uparrow}_i-z^{\uparrow}_j)^2
\prod_{k>l} (z^{\downarrow}_k-z^{\downarrow}_l)^2 \ .
\end{equation}
Using the Cauchy identity,
\begin{equation}
\label{eqn:Cauchy}
\prod^N_{i>j=1}(a_i-a_j)(b_i-b_j)=\prod^N_{i,j=1} (a_i-b_j)
{\rm Det}|(a_i-b_j)^{-1}| \ ,
\end{equation}
this can be rewritten as
\begin{eqnarray}
\Psi_{\rm pair} &=& \Psi_{(3,3,-1)} \cr
&=& \prod_{i>j} (z^{\uparrow}_i-z^{\uparrow}_j)^3
(z^{\downarrow}_i-z^{\downarrow}_j)^3
\prod_{i,j} (z^{\uparrow}_i-z^{\downarrow}_j)^{-1} \ .
\end{eqnarray}
Thus $\Psi_{\rm pair}$ is the $(3,3,-1)$
state if one takes the $p_x-ip_y$ pairing.
This wavefunction is well-behaved
in the long distance
limit, but has a short distance singularity.
In the presence of Landau-level mixing, the short distance
part of the wavefunction can be modified without changing
the structure of the wavefunction in the long distance limit.
\begin{eqnarray}
\label{eqn:(3,3,-1)}
\Psi_{(3,3,-1)} &=& {\rm Pf} \left [
h\left(\left|{z_i} - {z_j}\right|/\xi\right)\:
{\uparrow_i \downarrow_j +
\downarrow_i \uparrow_j \over z_i - z_j } \right ]\,\times
\cr & &{\hskip 1 cm}
\prod_{i>j} (z^{\uparrow}_i-z^{\uparrow}_j)^2
\prod_{k>l} (z^{\downarrow}_k-z^{\downarrow}_l)^2 \ .
\end{eqnarray}
Here, $h(0)=0$ and $h(x)\rightarrow 1$ as $x\rightarrow \infty$.
In realistic systems, where Landau-level mixing
is substantial, $\Psi_{(3,3,-1)}$ could be
a good candidate for the paired quantum Hall state represented
by $\Psi_{\rm pair}$.
It is natural to assume that $\Psi_{\rm pair}$
does not have an ``interlayer Josephson effect'' because
there is no gapless neutral mode in the system, in contrast
to the case of the $(1,1,1)$ state \cite{eisenstein,demler}.
We will show this later by direct calculation.

Another possibility for the pair wavefunction is an
exponentially-decaying function with a correlation length $\xi$,
for example either of
\begin{eqnarray}
\label{eqn:strong-pairing} \Psi_{\rm SP} &=& {\rm Pf} \left [
{e^{-|z_i-z_j|/\xi}}\: {\uparrow_i \downarrow_j + \downarrow_i
\uparrow_j \over z_i - z_j } \right ]\,\times \cr & &{\hskip 1 cm}
\prod_{i>j} (z^{\uparrow}_i-z^{\uparrow}_j)^2 \prod_{k>l}
(z^{\downarrow}_k-z^{\downarrow}_l)^2 \ ,
\cr \Psi_{\rm SP} &=&
{\rm Pf} \left [ {e^{-|z_i-z_j|/\xi}}\: \left(\uparrow_i
\downarrow_j - \downarrow_i \uparrow_j \right) \right ]\,\times
\cr & &{\hskip 1 cm} \prod_{i>j} (z^{\uparrow}_i-z^{\uparrow}_j)^2
\prod_{k>l} (z^{\downarrow}_k-z^{\downarrow}_l)^2 \ .
\end{eqnarray}
This would correspond to a ``strong'' pairing (SP) state  while
the previous choice  --- the $(3,3,-1)$ state --- corresponds to a
``weak'' pairing state in the terminology of Read and Green
\cite{green}. The two different choices of SP wavefunctions (with
$p$- and $s$-wave pairs, respectively) in
(\ref{eqn:strong-pairing}) can be continuously connected without
crossing a phase transition.

\section{$\lowercase{d} \stackrel{<}{\sim}  \ell_B$: $(1,1,1)$ state}

When the layer separation becomes sufficiently small, the
interlayer Coulomb interaction can be comparable to or even larger
than the intralayer Coulomb interaction. In this case, it should
be more advantageous to first form an ``interlayer'' dipolar
object which consists of an electron in one layer and two vortices
in the other layer, then form a paired state of these
``interlayer'' composite fermions, as shown in Fig.\ 3. In the
Chern-Simons formulation, this corresponds to the situation in
which the electron in one layer can only see fictitious flux in
the other layer. The appropriate Chern-Simons constraint equation
are:
\begin{equation}
\nabla \times {\bf a}_{\uparrow} = 2 \pi {\tilde \phi}
\delta \rho_{\downarrow} \ ,   \ \ \
\nabla \times {\bf a}_{\downarrow} = 2 \pi {\tilde \phi}
\delta \rho_{\uparrow} \ .
\end{equation}

As in the previous case, we can form symmetric and antisymmetric
combinations of the gauge fields ${\bf a}_{\uparrow}$
and ${\bf a}_{\downarrow}$. Again, the antisymmetric
combination mediates an attractive interaction.
The wavefunction of the corresponding paired quantum Hall state
has the following form.
\begin{equation}
\Phi_{\rm pair} =
\Phi^{\rm cf}_{\rm pair}
\prod^N_{i,j=1} (z^{\uparrow}_i-z^{\downarrow}_j)^2 \ ,
\end{equation}
where
\begin{equation}
\Phi^{\rm cf}_{\rm pair}
= {\rm Pf} [g(z_i, z_j ; \uparrow, \downarrow)]
\end{equation}
and $g(z_1, z_2 ; \uparrow, \downarrow)$ is the appropriate pair
wavefunction. Notice that the wavefunction $\Phi_{\rm pair}$ can
be regarded as the product of the wavefunction of a paired state
of the composite fermions and that of the (0,0,2) quantum Hall
state for bosons.

\begin{figure}
\center
\centerline{\epsfysize=1.4in
\epsfbox{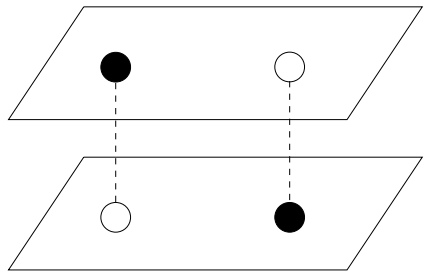}}
\begin{minipage}[t]{8.1cm}
\caption{The pairing of the ``interlayer''
composite fermions.
}
\label{fig3}
\end{minipage}
\end{figure}

However, this line of thinking appears to
conflict with the conventional wisdom
that a bilayer quantum Hall system
at $\nu_T=1$ is described by the $(1,1,1)$
state for $d/{\ell_B}\sim 1$ \cite{Halperin}.
Fortunately, the $(1,1,1)$ state,
\begin{equation}
\Psi_{(1,1,1)} = \prod^{N}_{i>j=1} (z^{\uparrow}_i-z^{\uparrow}_j)
(z^{\downarrow}_i-z^{\downarrow}_j)
\prod^{N}_{i,j=1} (z^{\uparrow}_i-z^{\downarrow}_j) \ .
\end{equation}
can be rewritten in the form
\begin{equation}
\Psi_{(1,1,1)} = {\rm Pf} \left [ {\uparrow_i \downarrow_j +
\downarrow_i \uparrow_j \over z_i - z_j } \right ]
\prod^{N}_{i,j=1} (z^{\uparrow}_i-z^{\downarrow}_j)^2 \ .
\end{equation}
using the Cauchy identity.
In other words,
\begin{equation}
g (z_i, z_j ; \uparrow, \downarrow) = {\uparrow_i \downarrow_j +
\downarrow_i \uparrow_j \over z_i - z_j } \
\end{equation}
is the correct choice for $d/{\ell_B}\sim 1$.

Notice that this form of $g (z_i, z_j ; \uparrow, \downarrow)$
corresponds to the (pseudo-)spin triplet $p_x - i p_y$ pairing
order parameter for the ``interlayer'' composite fermions. {}From
this point of view, it is natural to have the same pairing
symmetry (\ref{eqn:(3,3,-1)}) for $d>\ell_B$ but for
``intralayer'' rather than ``interlayer'' composite fermions.

\section{Phase Diagram at ${\nu_T}=1$: Chern-Simons Landau-Ginzburg
Description}

In this section, we consider these states within the framework of
Chern-Simons effective field theories
\cite{Zhang89,Read89,Read90,Blok90,Moon95}, consider the nature of
the transitions between them, and also find an additional state
which is a hierarchical descendent of the (3,3,-1) state,
described by a $3\times 3$ K-matrix.

The Lagrangian of the $(1,1,1)$ state is:
\begin{eqnarray}
{\cal L}^{(1,1,1)} &=& \Psi_{\uparrow}^{\dagger} ( i \partial_t +
A^0_{\uparrow} - a^0_{\uparrow} - a^0_{\downarrow} )
\Psi_{\uparrow}\cr & &\mbox{}-\frac{1}{2m} |(\frac{\bf \nabla}{i}
+ {\bf A}_{\uparrow} - {\bf a}_{\uparrow} - {\bf a}_{\downarrow} )
\Psi_{\uparrow}|^2\cr & &\mbox{}+ \frac{1}{4 \pi}
\epsilon^{\mu\nu\lambda} a^{\mu}_{\uparrow} \partial_\nu
a^{\lambda}_{\uparrow} +( \uparrow \rightarrow \downarrow )-
V_{int}. \label{cs-(1,1,1)}
\end{eqnarray}
Here $\Psi_{\uparrow\downarrow}$ and $A^\mu_{\uparrow\downarrow}$
describe composite bosons and electromagnetic fields in the two layers,
and the statistical gauge fields $a^\mu_{\uparrow\downarrow}$
ensure the agreement between this Lagrangian and the $(1,1,1)$
wavefunction. In the absence of interlayer
tunneling, the number of electrons in each layer is conserved, so
we can use the dual description \cite{duality}
of the $(1,1,1)$ state:
\begin{eqnarray}
{\cal L}_d^{(1,1,1)} &=& \frac{1}{2} |( i \partial_\mu - b_\mu^I)
\Phi_{vI}|^2 + V\left({\Phi_{vI}}\right) + \frac{1}{2}
(f_{\mu\nu}^I)^2\cr &
&\mbox{}+\frac{1}{4\pi}K_{IJ}b^I_{\mu}\partial_\nu b^J_{\lambda}
\epsilon^{\mu\nu\lambda} - \frac{1}{2\pi} A_\mu^a t^I_a
\partial_\nu b^I_{\lambda} \epsilon^{\mu\nu\lambda}. \label{Ld}
\end{eqnarray}
Here  $\Phi_{v I}$ describe vortices in the fields $\Psi_I$, with
indices $I$ and $J$ labeling the layers $\uparrow$ and
$\downarrow$, the dual gauge fields $b_\mu^I$ describe the
conserved currents, the Latin indices $\mu$, $\nu$, $\lambda$
include space and time components, and the Gram (or K-)matrix
\cite{Read90,Blok90} is
\begin{eqnarray}
\label{eqn:K-(1,1,1)}
K_{IJ} = \left( \begin{array}{cc} 1 & 1 \\ 1 & 1 \end{array}
\right).
\end{eqnarray}

It is also convenient to define the charge and spin gauge fields
$A^{C,S}_\mu = (A_\mu^{\uparrow} \pm A_\mu^{\downarrow})/2$,
$b^{C,S}_\mu = b_\mu^{\uparrow} \pm b_\mu^{\downarrow}$ with
charge and spin vectors ${\bf t}_C=(1,1)$ and ${\bf t}_S=(1,-1)$,
and $a=C$, $S$. Here and henceforth, we use the term `spin' to
refer to the charge difference between the two layers, not the
physical spin of the electrons, which is assumed to be fully
polarized (i.e.\ spin here refers to an pseudospin associated with
the layers).

A generic quasiparticle may now be constructed by taking a
composite of $l_1$ vortices of type $\Phi_{v \uparrow}$ and $l_2$
vortices of type $\Phi_{v \downarrow}$. $\Phi_{({l_1},{l_2})}$
creates such a quasiparticle, which has charge $Q$ and
spin $S$
\begin{eqnarray}
\label{eqn:charge,spin}
Q &=& t_C^{T} K^{-1} l, \cr S &=& t_S^{T} K^{-1} l.
\end{eqnarray}
If the $K$-matrix has vanishing determinant, as it does in
(\ref{eqn:K-(1,1,1)}), then equation (\ref{eqn:charge,spin}) will
have to be modified. When this occurs, the zero eigenvalue
corresponds to the Goldstone modes associated with some broken
continuous symmetry. Hence, if we are to use the K-matrix
formalism to calculate the quasiparticle properties and the
degeneracy of the ground states on a torus (usually the degeneracy
is $\det K$ \cite{Read90,Blok90}), some sort of reduced K-matrix
will be required. We will now describe how this can be done in
general. It is helpful to think in terms of the vectors in the
condensate lattice \cite{Read90}; the K-matrix is the Gram matrix
of the lattice, that is the matrix of inner products of a basis of
vectors in the lattice. The inner product of two vectors $\bf m$,
$\bf n$ in the lattice, represented as column vectors of integers
[not to be confused with the similar vectors $l$, which lie
instead in the dual (excitation) lattice], is then given by ${\bf
m}^T K{\bf n}$. The vanishing determinant of $K$ implies that we
can find a lattice vector $\bf n$ such that $K{\bf n}={\bf 0}$.
Then the inner product of $\bf n$ with {\em any} other vector,
including itself, is zero; we call $\bf n$ a {\em null vector}. We
choose $\bf n$ to be {\em primitive}, that is, not divisible by
any integer larger than $1$. Two vectors that differ by an integer
multiple of $\bf n$ have the same inner product with any other
vector, because $\bf n$ is null. Hence, we can obtain a reduced
lattice, in which we identify vectors that differ by integer
multiples of $\bf n$, and the inner product remains well-defined.
The reduced lattice is the quotient of the previous one by $\bf
n$. In terms of matrices, the reduced K-matrix is obtained by
changing basis \cite{Read90,Blok90}, taking $\bf n$ as one of the
basis vectors. Then in the resulting K-matrix, the entries in the
row and column corresponding to $\bf n$ are all zero. The reduced
K-matrix, $K_{\rm red}$ is obtained by deleting this row and
column. The process can be repeated until either a nonzero
determinant is obtained, or the lattice has dimension zero; in our
examples, a single reduction is sufficient.

In the case of the $(1,1,1)$ state above, the null vector is ${\bf
n}= (1,-1)$, and the reduced K-matrix is $K_{{\rm red}\,IJ}=(1)$,
the same as in the polarized or single-layer $\nu=1$ state. The
physical meaning of the procedure is that the $(1,1,1)$ state is
obtained by condensing composite bosons with pseudospin, or by
taking a pseudospin-polarized state and tilting the pseudospins
into the XY plane. This does not affect the quantum Hall
properties of the state, which remain those of the single layer
$\nu=1$ state. The procedure above correctly accounts for
disregarding the direction of the pseudospin, and implies that
there is a single ground state on the torus, up to low-lying
states associated with the broken symmetry. The use of the reduced
K-matrix gives the quasiparticle properties; the quasiparticles
carry charge $\pm 1$ and are fermions. We see that the merons
(vortices in the pseudospin order parameter that carry charge $\pm
1/2$ and ill-defined statistics) are not obtained from $K_{\rm
red}$, but they are confined by the logarithmic potential between
them, and cannot be separated to infinity with finite energy.
Usually, the different degenerate ground states can be obtained
from each other by creating a quasiparticle-quasihole pair,
transporting one of them around the torus, and subsequently
annihilating them. The nontrivial statistics (abelian, in all
cases in this paper) of the quasiparticles then require degenerate
ground states. Since the merons are confined, they do not
contribute to the count of ground states, and indeed, dragging one
around the torus produces a helical texture in the ground state,
increasing the energy by order width/length; we do not regard such
a state as a ground state. In general, the ground-state degeneracy
is divisible by the denominator of the filling factor $\nu_T$
(equal to $1$ here); any ground-state degeneracy beyond that is
not exact in a finite size system, but the energy splitting $\sim
\exp(-c L)$ on a torus of size $L$, where $c$ is a constant.
Finally, spin wave states have excitation energies $\sim 1/L$.

Returning to the dual Lagrangian in terms of the unreduced
K-matrix of the $(1,1,1)$ state, in terms of the charge and spin
gauge fields and the quasiparticle fields $\Phi_{(m,n)}$, it is:
\begin{eqnarray}
{\cal L}_d^{(1,1,1)} &=& \frac{1}{2} \biggl|\biggl( i \partial_\mu
- \left(\frac{m+n}{2}\right)\,b^C_\mu -\cr & & {\hskip 3 cm}
\left(\frac{m-n}{2}\right)\,b^S_\mu \biggr)
\Phi_{(m,n)}\biggr|^2\cr & &\mbox{}+\frac{1}{4\pi}
b^C_{\mu}\partial_\nu b^C_{\lambda} \epsilon^{\mu\nu\lambda} -
\frac{1}{2\pi} A_\mu^a
\partial_\nu b^a_{\lambda} \epsilon^{\mu\nu\lambda}\cr
&&\mbox{} +\frac{1}{2} (f_{\mu\nu}^a)^2.\!\!\!
\label{Ld-c-s-(1,1,1)}
\end{eqnarray}

Since there is no Chern-Simons term for $b^S_\mu$, it is massless.
This gauge field is dual to the Goldstone mode which results when
$\Psi_\uparrow$, $\Psi_\downarrow$ condense, thereby breaking the
$U(1)$ pseudospin symmetry. Quantum fluctuations can disorder the
pseudospin degree of freedom. This occurs when $\Phi_{(1,-1)}$
(the field for merons) condenses in eq.\ (\ref{Ld-c-s-(1,1,1)}).
The effective theory for this transition is:
\begin{eqnarray}
{\cal L}_d^{(1,1,1)} &=& \frac{1}{2} \left|\left( i \partial_\mu -
\,b^S_\mu \right) \Phi_{(1,-1)}\right|^2 +
V\left({|{\Phi_{(1,-1)}}|^2}\right)\cr & &\mbox{}+\frac{1}{4\pi}
b^C_{\mu}\partial_\nu b^C_{\lambda} \epsilon^{\mu\nu\lambda} -
\frac{1}{2\pi} A_\mu^a  \partial_\nu b^a_{\lambda}
\epsilon^{\mu\nu\lambda}\cr&&\mbox{} +\frac{1}{2}
(f_{\mu\nu}^a)^2.
\end{eqnarray}
Applying $U(1)$ duality \cite{duality} in reverse to the
pseudospin alone, we find
\begin{eqnarray}
{\cal L}_d^{(1,1,1)} &=& \frac{1}{2} \left|
\left(i\partial_\mu-{A_\mu^S}\right) \phi
\right|^2 + V\left({|\phi|^2}\right)\cr & &\mbox{}+\frac{1}{4\pi}
b^C_{\mu}\partial_\nu b^C_{\lambda} \epsilon^{\mu\nu\lambda} -
\frac{1}{2\pi} A_\mu^C  \partial_\nu b^C_{\lambda}
\epsilon^{\mu\nu\lambda}\cr &&\mbox{} +\frac{1}{2}
(f_{\mu\nu}^a)^2. \label{(1,1,1)-SP}
\end{eqnarray}
The first line of eq.\ (\ref{(1,1,1)-SP}) decouples from the
second so the transition between the $(1,1,1)$ state and the
quantum disordered state is in the XY universality class.

According to the arguments of \cite{demler}, such a disordered
state is the SP state of (\ref{eqn:strong-pairing}), with
Landau-Ginzburg theory:
\begin{eqnarray}
{\cal L}_p &=& \Psi_p^{\dagger} ( \partial_0 - a_0 - 2 A_0^{C} )
\Psi_p \cr & &\mbox{} + \,\frac{1}{2 m} \left| \left[
\vec{\partial} - i \vec{a} - 2i \vec{A}^{C} \right] \Psi_p
\right|^2\cr &&\mbox{}
 - \, \frac{1}{16 \pi }\, \epsilon_{\mu\nu\lambda}
a_{\mu} \partial_{\nu} a_{\lambda}. \label{Lp}
\end{eqnarray}
The order parameter is given by ${\Psi_p}={\Psi_\uparrow}
{\Psi_\downarrow}$ (as opposed to ${\Psi_{\uparrow,\downarrow}}$
individually as in eq.\ (\ref{cs-(1,1,1)})). Equation (\ref{Lp})
may be derived from (\ref{cs-(1,1,1)}) in the limit that
${\Psi_p}$ is lighter than ${\Psi_{\uparrow,\downarrow}}$. The
dual theory for SP is:
\begin{eqnarray}
{\cal L}_d^{SP} &=& \frac{1}{2} \left|\left( i \partial_\mu -
\,b^C_\mu \right) \Phi_{l}\right|^2\cr & &\mbox{}+\frac{4}{4\pi}
b^C_{\mu}\partial_\nu b^C_{\lambda} \epsilon^{\mu\nu\lambda} - 2
\frac{1}{2\pi} A_\mu^C  \partial_\nu b^C_{\lambda}
\epsilon^{\mu\nu\lambda}\cr &&\mbox{} +\frac{1}{2}
(f_{\mu\nu}^C)^2. \label{Ld-SP}
\end{eqnarray}
According to standard arguments \cite{Read90,Blok90}, the SP
ground states on a torus are $4$-fold degenerate, and these ground
states all have even electron number \cite{green}.

We note the existence, in principle, of a state intermediate
between the SP and $(1,1,1)$ states with the following
Landau-Ginzburg theory
\begin{eqnarray}
{\cal L}_p &=& \Psi_p^{\dagger} ( \partial_0 - a_0 - 2 A_0^{C} )
\Psi_p \cr & &\mbox{} + \,\frac{1}{2 m} \left| \left[
\vec{\partial} - i \vec{a} - 2i \vec{A}^{C} \right] \Psi_p
\right|^2
 - \, \frac{i}{16 \pi }\, \epsilon_{\mu\nu\lambda}
a_{\mu} \partial_{\nu} a_{\lambda}\cr & &\mbox{} +
\Psi_s^{\dagger} (
\partial_0 - 2 A_0^{S} ) \Psi_s + \,\frac{1}{2 m} \left| \left[
\vec{\partial}
 - 2i \vec{A}^{S} \right] \Psi_s \right|^2\cr
& & \mbox{}+ V\left({|\Psi_p|^2}, {|\Psi_s|^2}\right),
\label{Lp,s}
\end{eqnarray}
which is valid when ${\Psi_p}={\Psi_\uparrow} {\Psi_\downarrow}$
and ${\Psi_s}={\Psi_\uparrow} {\Psi_\downarrow^\dagger}$ are light
fields. The state in which $\langle {\Psi_p}\rangle \neq  0$,
$\langle {\Psi_s}\rangle \neq  0$ also breaks pseudospin symmetry,
as a result of the latter order parameter. We will call this state
SP/F to indicate the coexistence of distinct strong pairing and
ferromagnetic order parameters. Note that again the K-matrix for
the SP state, and the reduced K-matrix for the SP/F state, are
both just $K_{IJ}=(4)$. The transition between the SP and SP/F
states is an $XY$ transition at which $\Psi_s$ condenses. The
transition between the SP/F and $(1,1,1)$ states is an Ising
transition at which the symmetry ${\Psi_{\uparrow,\downarrow}}
\rightarrow - {\Psi_{\uparrow,\downarrow}}$ is broken; it can also
be viewed as a strong- to weak-pairing transition, similar to
Ref.\ \cite{green}, but in the presence of ferromagnetic order in
the pseudospin.

An alternative Chern-Simons Landau-Ginzburg theory yields the
$(3,3,-1)$ state, by Bose-condensing composite bosons that consist
of an up electron plus 3 vortices acting on the up electrons, and
-1 vortices acting on the down electrons:
\begin{eqnarray}
{\cal L}^{(3,3,-1)} &=& \Psi_{\uparrow}^{\dagger} ( i \partial_t +
A^0_{\uparrow} - 3 a^0_{\uparrow} + a^0_{\downarrow} )
\Psi_{\uparrow}\cr & &\mbox{}-\frac{1}{2m} |(\frac{\bf \nabla}{i}
+ {\bf A}_{\uparrow} - 3 {\bf a}_{\uparrow} + {\bf a}_{\downarrow}
) \Psi_{\uparrow}|^2\cr & &\mbox{}+ \frac{1}{4 \pi}
\epsilon^{\mu\nu\lambda} a^{\mu}_{\uparrow} \partial_\nu
a^{\lambda}_{\uparrow} +( \uparrow \rightarrow \downarrow )-
V_{int}. \label{cs-(3,3,-1)}
\end{eqnarray}
As we have described in the previous section, this state can also
be viewed as a paired state. Passing to the dual theory, we have
eq.\ (\ref{Ld}), but with
\begin{eqnarray}
K_{IJ} = \left( \begin{array}{rr} 3 & -1 \\ -1 & 3 \end{array}
\right).
\label{K33-1}
\end{eqnarray}

In terms of the charge and spin gauge fields
and the quasiparticle fields $\Phi_{(m,n)}$,
the dual Lagrangian takes the form:
\begin{eqnarray}
{\cal L}_d^{(3,3,-1)} &=& {\cal L}\left({\Phi_{(m,n)}}\right)\cr &
& \mbox{}+\frac{1}{4\pi} b^C_{\mu}\partial_\nu b^C_{\lambda}
\epsilon^{\mu\nu\lambda} +\frac{2}{4\pi} b^S_{\mu}\partial_\nu
b^S_{\lambda} \epsilon^{\mu\nu\lambda}\cr & &\mbox{} -
\frac{1}{2\pi} A_\mu^a  \partial_\nu b^a_{\lambda}
\epsilon^{\mu\nu\lambda} +\frac{1}{2} (f_{\mu\nu}^a)^2.
\label{Ld-c-s-(3,3,-1)}
\end{eqnarray}
{}From (\ref{K33-1}) we can see that the charge sector
of the (33-1) state is similar to the SP state. It has a quantized
charge Hall conductance and supports
elementary excitations of charge 1/2.
The pseudospin sector is different from that of either of the
other states: it is gapped --- unlike $(1,1,1)$ --- and exhibits a
pseudospin Hall effect (which is manifested in the Hall drag
resistance, as we discuss later), unlike $(1,1,1)$ and SP.

The condensation of the neutral semion $(1,-1)$ in
(\ref{Ld-c-s-(3,3,-1)}) eliminates the peudospin gauge field
$b^S_\mu$ by the Anderson-Higgs effect, thereby leading to the SP
state. This is analogous to the situation at $\nu_T=1/2$, where it
was shown in \cite{green} that the transition between the
$(3,3,1)$ and strong-pairing states is a second-order transition
at which a Dirac fermion becomes massless. However, in the
${\nu_T}=1$ case, we are dealing with a semion, rather than a
fermion, so we might expect the transition to be analogous to the
quantum Hall liquid to insulator transition. Both are described by
a single relativistic field coupled to the Chern-Simons gauge
field. In the large $N$ limit this transition was shown to be
second order \cite{Wen93,Wen99}, but in the relevant $N=1$ limit
the gauge field fluctuations may drive the transition first order
\cite{Pryadko94}. However, similar arguments in the absence of a
Chern-Simons term were not conclusive. That transition was argued
to be first-order in $4-\epsilon$ dimensions \cite{Coleman73}, but
$3D$ duality \cite{duality} implies that the transition is in the
inverted $XY$ universality class and, therefore, second order.
Therefore, the possibility that the transition in the presence of
the Chern-Simons term is second order appears to be still open. In
the presence of disorder, at any rate, the transition will be
second order. We believe, therefore, that the ${\nu_T} =1$ bilayer
quantum phase transition between the $(3,3,-1)$ paired state and
the SP state is a continuous phase transition.

We may, on the other hand, consider the
condensation of the boson $(2,-2)$ upon
the attachment of $2$ flux quanta:
\begin{eqnarray}
\label{eqn:(2,-2)} {\cal L}_d^{(3,3,-1)} &=& \frac{1}{2}
\left|\left( i \partial_\mu - 2 b^S_\mu - {\alpha_\mu} \right)
\tilde{\Phi}_{(2,-2)}\right|^2 + V\left( |\tilde{\Phi}_{2,-2}|^2
\right)\cr & & \mbox{}+ \frac{1}{4\pi} \alpha_\mu
\partial_\nu \beta_\lambda \epsilon^{\mu\nu\lambda}
+ \frac{2}{4\pi} \beta_\mu
\partial_\nu \beta_\lambda \epsilon^{\mu\nu\lambda} \cr
& &\mbox{} +\frac{1}{4\pi} b^C_{\mu}\partial_\nu b^C_{\lambda}
\epsilon^{\mu\nu\lambda} +\frac{2}{4\pi} b^S_{\mu}\partial_\nu
b^S_{\lambda} \epsilon^{\mu\nu\lambda}\cr & &\mbox{} -
\frac{1}{2\pi} A_\mu^a  \partial_\nu b^a_{\lambda}
\epsilon^{\mu\nu\lambda} +\frac{1}{2} (f_{\mu\nu}^a)^2.
\end{eqnarray}
When $\tilde{\Phi}_{2,-2}$ condenses, the resulting Meissner
effect enforces the condition $2 b^S_\mu = {\alpha_\mu}$ (up to
gauge transformations). Hence, the following quantum Hall state
results:
\begin{eqnarray}
{\cal L} &=&
\frac{4}{4\pi} b^S_\mu
\partial_\nu \beta_\lambda \epsilon^{\mu\nu\lambda}
+ \frac{2}{4\pi} \beta_\mu
\partial_\nu \beta_\lambda \epsilon^{\mu\nu\lambda} \cr
& &\mbox{} +\frac{1}{4\pi} b^C_{\mu}\partial_\nu b^C_{\lambda}
\epsilon^{\mu\nu\lambda} +\frac{2}{4\pi} b^S_{\mu}\partial_\nu
b^S_{\lambda} \epsilon^{\mu\nu\lambda}\cr & &\mbox{} -
\frac{1}{2\pi} A_\mu^a  \partial_\nu b^a_{\lambda}
\epsilon^{\mu\nu\lambda} +\frac{1}{2} (f_{\mu\nu}^a)^2,
\end{eqnarray}
or,
\begin{eqnarray}
\label{eqn:descendent} {\cal L} &=& \frac{1}{4\pi}
\,{K_{IJ}}\,b^I_{\mu}\partial_\nu b^J_{\lambda}
\epsilon^{\mu\nu\lambda} + \frac{1}{2} (f_{\mu\nu}^I)^2\cr & &
\mbox{}- \frac{1}{2\pi} A_\mu^a \, {t^I_a}\,\partial_\nu
b^I_{\lambda} \epsilon^{\mu\nu\lambda},
\end{eqnarray}
where $I,J=1,2,3$, $b^3_\mu\equiv \beta_\mu$,
${t^I_C}=(1,1,0)$, ${t^I_S}=(1,-1,0)$, and
\begin{eqnarray}
K_{IJ} = \left( \begin{array}{rrr} 3 & -1 & 2 \\ -1 & 3 & -2
\\ 2 & -2 & 2 \end{array}
\right).
\end{eqnarray}
This state is a hierarchical descendent of the $(3,3,-1)$ state
(though the construction differs from the usual hierarchy by
condensing neutral quasiparticles, which does not change the
filling factor). A wavefunction for it can be constructed along
the lines of Ref.\ \cite{Read90}. When the number of flux
quanta on the sphere is $N_\phi=N-3$ (as in the $(3,3,-1)$ state),
it contains two merons and two anti-merons,
which are bound in pairs to form two
charge $+1$ excitations, so the natural ground state has
$N_\phi=N-1$. Also, it can occur for $N$ odd as well as for $N$
even, like the $(1,1,1)$ state, and unlike the $(3,3,-1)$ state.
The state is distinct from the $(1,1,1)$ state, despite the fact
that it breaks pseudospin symmetry (since $\tilde{\Phi}_{(2,-2)}$
carries pseudospin, the same value as the electron) and has a
gapless Goldstone mode.
The $3\times 3$ K-matrix has
determinant zero, and hence a reduced K-matrix is required. This
can be obtained most easily by first making the basis change for
the condensate lattice to basis vectors $(1,0,-1)$, $(0,1,1)$, and
$(0,0,1)$ (relative to the previous basis). The resulting K-matrix
is
\begin{eqnarray}
K'_{IJ} = \left( \begin{array}{rrr} 1 & 1 & 0 \\ 1 & 1 & 0
\\ 0 & 0 & 2 \end{array}
\right).\label{K_3new}
\end{eqnarray}
Since this contains the $(1,1,1)$ state K-matrix as a block, it is
clear that the reduced K-matrix is
\begin{eqnarray}
K'_{{\rm red}\,IJ} = \left( \begin{array}{rr} 1 & 0 \\ 0 & 2
\end{array} \right).\label{Krednew}
\end{eqnarray}
Hence, the ground state degeneracy on the torus is $2$.
Incidentally, the block containing only $(2)$ represents a
``hidden SU($2$)'' in this state; the corresponding edge theory is
an SU($2$) current algebra at level $1$, even though this will
presumably not be a symmetry of the Hamiltonian.

We can complete the circle and return to our starting point, the
$(1,1,1)$ state, if the quasiparticle $\Phi_{(1,-1,1)}$ (where the
vortices are relative to the original basis) condenses in the
state (\ref{eqn:descendent}), thereby eliminating one of the
neutral gauge fields by the Anderson-Higgs effect. The
proliferation of these vortices leaves intact only those
condensates (composite boson fields) which do not wind on going
around these vortices. These condensates lie on a sublattice (of
the unreduced lattice), which is the same as that of the $(1,1,1)$
state; in fact in the basis used for $K'$ in eq.\ (\ref{K_3new}),
or for $K'_{\rm red}$ in eq.\ (\ref{Krednew}), the transition has
destroyed the condensate described by the $1\times 1$ block at the
lower right, leaving the $(1,1,1)$ state. This transition could be
first-order or second order in the absence of disorder, according
to the conflicting conventional wisdom discussed above.

A more useful form for the critical theory for the transition
between $(3,3,-1)$ and its interlayer coherent descendent, along
the lines of eq.\ (\ref{(1,1,1)-SP}), may be derived from eq.\
(\ref{eqn:(2,-2)}) by making the change of variables
${\alpha_\mu}\rightarrow {\alpha_\mu} - 2 b_\mu^S$ and integrating
out $\beta_\mu$. The Lagrangian takes the form
\begin{eqnarray}
{\cal L} &=& \frac{1}{2} |( i \partial_\mu -\alpha_\mu )
\tilde{\Phi}_{2,-2} |^2 + V \nonumber\\ & &\mbox{} -\frac{1}{8\pi}
\alpha_\mu
\partial_\nu \alpha_\lambda \epsilon^{\mu\nu\lambda}
+ \frac{1}{2\pi} \alpha_\mu
\partial_\nu b_\lambda^S
\epsilon^{\mu\nu\lambda} \nonumber\\ & &\mbox{}+\frac{1}{2}
(f_{\mu\nu}^a)^2 +\frac{1}{4\pi} b^C_{\mu}\partial_\nu
b^C_{\lambda} \epsilon^{\mu\nu\lambda} -\frac{1}{2\pi} A_\mu^a
\partial_\nu b^a_{\lambda} \epsilon^{\mu\nu\lambda}.
\end{eqnarray}
The gauge field $b_\mu^S$ only appears linearly in the Lagrangian,
so we may integrate it out, thereby resulting
in the constraint that $\alpha_\mu=A_\mu^S$ up to
a gauge transformation. The final Lagrangian is
then
\begin{eqnarray}
\label{eqn:Hall_S} {\cal L} &=& \frac{1}{2} |( i \partial_\mu
-A_\mu^S ) \tilde{\Phi}_{2,-2} |^2 +
V\left(|\tilde{\Phi}_{2,-2}|^2 \right)\nonumber\\ & &\mbox{}
-\frac{1}{8\pi} {A^S_\mu}
\partial_\nu {A^S_\lambda} \epsilon^{\mu\nu\lambda}\nonumber\\
& &\mbox{}+ \frac{1}{2} (f_{\mu\nu}^I)^2
+\frac{1}{4\pi}b^C_{\mu}\partial_\nu b^C_{\lambda}
\epsilon^{\mu\nu\lambda} - \frac{1}{2\pi} A_\mu^C \partial_\nu
b^C_{\lambda} \epsilon^{\mu\nu\lambda}.
\end{eqnarray}
Hence, the transition between the $(3,3,-1)$
state and its interlayer coherent hierarchical descendent
is also in the XY universality class.

\begin{figure}
\vspace{-0.5truecm} \center \centerline{\epsfysize=2.3in
\epsfbox{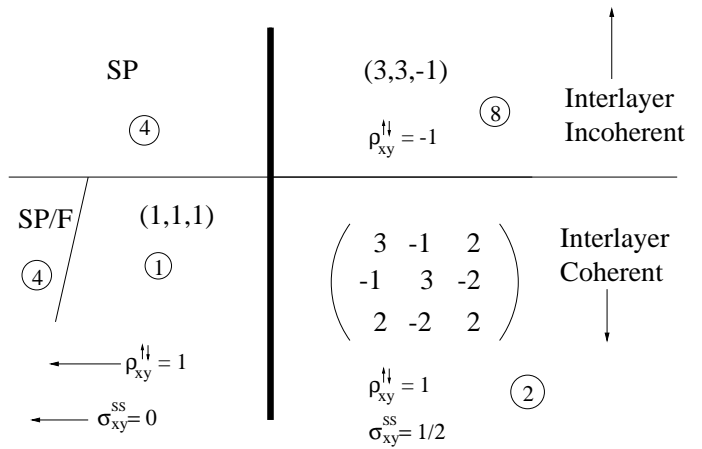}}
\begin{minipage}[t]{8.1cm}
\caption{Schematic phase diagram of states at ${\nu_T}=1$. The
thick line represents a phase transition which may be first- or
second-order (see text). The horizontal thin line represents a
second-order phase transition in the XY universality class. The
thin line separating the SP/F and $(1,1,1)$ states is in the Ising
universality class. The relationship between the states in this
phase diagram is discussed in section V. The drag resistivity --
which, together with interlayer tunneling, can distinguish these
states -- is discussed in section VI. The ground state
degeneracies on the torus are encircled.} \label{fig4}
\end{minipage}
\end{figure}

These states and transitions are depicted in the phase diagram of
Figure \ref{fig4}. The two states in the upper portion of the
phase diagram --- SP and $(3,3,-1)$ --- are not interlayer
coherent while the three states in the lower portion [SP/F,
$(1,1,1)$, and the interlayer coherent descendent of $(3,3,-1)$
(identified by its $K$-matrix)] are interlayer coherent (i.e. they
spontaneously break pseudospin $U(1)$ symmetry). The development
of interlayer coherence may be probed by interlayer tunneling
experiments \cite{eisenstein}. The states on the right in Figure
\ref{fig4} are expected for $d\gg {\ell_B}$ while those on the
left are expected for $d\sim {\ell_B}$. As we discuss in the next
section, these may be distinguished by their Hall drag
resistivities. This phase diagram suggests that the transition
between the $(3,3,-1)$ and $(1,1,1)$ states may occur via an
intermediate state which is either the SP state or the interlayer
coherent hierarchical descendent of the $(3,3,-1)$ state. We
caution the reader that first-order transitions between any of
these states are possible, even those that are not adjacent in the
Figure. However, the $XY$ transitions {\it can} be second order.

\section{Drag Resistivities}

It is important to discuss experiments
which can distinguish the
$\Psi_{(1,1,1)}$ state, the $\Psi_{(3,3,-1)}$,
its interlayer coherent descendent, and the SP state.
Here, we propose Coulomb drag experiments in which
the longitudinal, $\rho^{\uparrow \downarrow}_{xx}$ and
Hall, $\rho^{\uparrow \downarrow}_{xy}$, drag resistivities
are used to distinguish the different
phases. These may be calculated in Chern-Simons
theory from
\begin{equation}
\label{eqn:drag}
{\rho_{ij}^{\alpha\beta}} = {\rho_{ij}^{{\rm cf}\:\alpha\beta}}
+ {\epsilon_{ij}}\: {\rho^{{\rm cs}\:\alpha\beta}}
\end{equation}
where $i,j=x,y$, $\alpha,\beta=\uparrow,\downarrow$,
$\epsilon_{ij}$ is the antisymmetric tensor,
and ${\rho^{{\rm cs}\:\alpha\beta}}=2{\delta_{\alpha\beta}}$
in the compressible and $(3,3,-1)$ states
(``intralayer'' composite fermions)
while ${\rho^{{\rm cs}\:\alpha\beta}}=2{\sigma^x_{\alpha\beta}}$
in the $(1,1,1)$ state (``interlayer'' composite fermions).

First, consider the longitudinal drag resistivity.
In the compressible state, if we neglect gauge field
fluctuations, $\rho_{xx}^{{\rm cf}\:\uparrow \downarrow}=0$,
so $\rho_{xx}^{\uparrow \downarrow}=0$. Including
these fluctuations,
it vanishes as $T^{4/3}$ at low temperatures \cite{kim}.
In $\Psi_{(1,1,1)}$ and $\Psi_{(3,3,-1)}$
(as well as its interlayer coherent descendent),
$\rho_{xx}^{{\rm cf}\:\uparrow \downarrow}$ and, hence,
$\rho_{xx}^{\uparrow \downarrow}$
vanish at zero temperature and are activated
at low temperatures \cite{vignale}.

Now, let us consider the Hall drag resistivity.
In the compressible state, both terms on
the right-hand-side of (\ref{eqn:drag})
vanish, so the Hall drag resistivity vanishes.
In the $(3,3,-1)$ state, ${\rho^{\rm cf}_{xy}}$
is that of a $p_x + i p_y$ superconductor,
which has vanishing charge resistivity
(since it is a superconductor) but quantized
spin Hall resistivity \cite{green,senthil}. In other words,
${\rho_{xy}^{{\rm cf}\: cc}}=0$,
${\rho_{xy}^{{\rm cf}\: cs}}=0$,
${\rho_{xy}^{{\rm cf}\: ss}}=1$.
Consequently, ${\rho_{xy}^{\uparrow \uparrow}}=
{\rho_{xy}^{\downarrow \downarrow}}=3$,
and ${\rho_{xy}^{\uparrow \downarrow}}=-1$.

In the $(1,1,1)$ state, ${\rho^{\rm cf}_{xy}}$
is identical, but ${\rho^{\rm cs}_{xy}}$
is different, so ${\rho_{xy}^{\uparrow \uparrow}}=
{\rho_{xy}^{\downarrow \downarrow}}=1$,
and ${\rho_{xy}^{\uparrow \downarrow}}=1$,
in agreement with \cite{yang}. The same result can be
deduced more physically by noting that interlayer
coherence requires that the voltage be the same
in both layers. If we run a current in one layer
alone, then this condition can only be
satisfied if ${\rho_{xy}^{\uparrow \uparrow}}=
{\rho_{xy}^{\uparrow \downarrow}}$. On the other
hand, the total Hall resistance of the system is
$({\rho_{xy}^{\uparrow \uparrow}}+{\rho_{xy}^{\uparrow \downarrow}})/2$.
Since this must equal $1$, we obtain the
previously stated result. Note that the
same logic applies to the interlayer
coherent descendent of the $(3,3,-1)$ state
which must, therefore, have ${\rho_{xy}^{\uparrow \uparrow}}=
{\rho_{xy}^{\uparrow \downarrow}}=1$. In other words,
the full resistivity tensor of the $(1,1,1)$
state is identical to that of the interlayer
coherent descendent of the $(3,3,-1)$ state.

Note that the interlayer coherent descendent of the $(3,3,-1)$
state discussed in the previous section is a {\it pseudospin Hall
superconductor}. {}From (\ref{eqn:Hall_S}), we see that the
pseudospin conductivity tensor is of the form
\begin{eqnarray}
\label{eqn:spin_Hall}
{\sigma^{SS}} = \left( \begin{array}{cc} \frac{\kappa}{i\omega}
& \frac{1}{2}  \\ -\frac{1}{2} & \frac{\kappa}{i\omega}
\end{array}
\right),
\end{eqnarray}
where $\kappa$ is a constant.
Hence, there is non-vanishing spin Hall conductivity.
However, upon inverting this tensor, we see that
the spin Hall resistivity
vanishes, as it must in order to satisfy
${\rho_{xy}^{\uparrow \uparrow}}=
{\rho_{xy}^{\uparrow \downarrow}}=1$.
The $(1,1,1)$ state, on the other hand,
has vanishing spin Hall conductivity.
The distinction between a {\it pseudospin Hall superconductor}
and an `ordinary' pseudospin superconductor is
reminiscent of the difference between a Hall insulator
and an ordinary insulator (but inverted).

Thus far, we have focused on the situation in which the layers are
perfectly balanced. If they are unbalanced due to the presence of
an external bias field, for example, this is analogous to
introducing a pseudospin Zeeman field along the $z$-direction.
This will have a pair-breaking effect on the paired states and
will be expected to weaken the quantum Hall effect. This should,
as a consequence, increase the longitudinal drag (along with the
total longitudinal resistance). The presence of external bias can,
thus, be used to distinguish the paired state from other
incompressible states.

We conclude by summarizing our results. We
have shown that the $\nu={1\over 2}$ (${\nu_T}=1$)
bilayer quantum Hall system (in the absence of interlayer
tunneling) is likely to have as its ground state a
novel paired Hall state (possibly of $p$-wave symmetry)
for intermediate layer separations $d>{\ell_B}$,
which gives way to the usual $(1,1,1)$ state
for smaller layer separations
($d \stackrel{<}{\sim} {\ell_B}$),
and to compressible Fermi-liquid-type states (two
decoupled Halperin-Lee-Read \cite{HLR}
$\nu={1\over 2}$ layers) for large layer separations
($d\gg{\ell_B}$). We argue that the quantum phase transitions
separating the paired states from the $(1,1,1)$ and
bilayer Halperin-Lee-Read states can be experimentally
studied via the measurement of various components
of interlayer drag resistivities. We also argue
that the transition between
the $(3,3,-1)$ and $(1,1,1)$ states may occur
via an intermediate state which is either
the SP state or the interlayer coherent hierarchical
descendent of the $(3,3,-1)$ state, and in either
case one of the two transitions will be in
the XY universality class.

\begin{acknowledgements}
We would like to thank L. Balents, J.
Eisenstein, and S. Sachdev
for discussions.
YBK, CN, NR, and SDS would like to thank
the Aspen Center for Physics for hospitality.
This work was supported by the NSF
under grant numbers DMR-9983783 (YBK),
DMR-9983544 (CN), DMR-9818259 (NR);
the A.P. Sloan Foundation (YBK and CN);
the Harvard Society of Fellows (ED); and the
ONR (SDS).
\end{acknowledgements}

\end{multicols}


\begin{references}


\bibitem{footnote_1}
Throughout this paper, we use $\nu_T$ to denote the total
filling factor and $\nu$ to denote the filling
factor in each layer, ${\nu_T}=2\nu$.


\bibitem{Eisenstein97}
See, for example, the articles by J.~P. Eisenstein,
S.~M. Girvin and A.~H. MacDonald in {\it Perspectives
in Quantum Hall Effects}, edited by S. Das Sarma
and A. Pinczuk (Wiley, New York, 1997); and references therein.


\bibitem{Halperin} B. I. Halperin, Helv. Phys. Acta {\bf 56}, 75 (1983);
Surf. Sci. {\bf 305},1 (1994).


\bibitem{eisenstein}
I. B. Spielman, J. P. Eisenstein, L. N. Pfeiffer, and K. W. West,
cond-mat/0002387.


\bibitem{Wen92}
X.~G. Wen and A. Zee, Phys. Rev. Lett.
{\bf 69}, 1811 (1992); Phys. Rev. B {\bf 47},
2265 (1993).

\bibitem{Zhang89}
S.C. Zhang {\it et al},  Phys. Rev. Lett. {\bf 62}, 82 (1989).

\bibitem{Read89}
N. Read,  Phys. Rev. Lett. {\bf 62}, 86 (1989).

\bibitem{Read90} N. Read, \prl {\bf 65}, 1502 (1990).

\bibitem{Blok90} B. Blok and X.-G. Wen, \prb {\bf 42}, 8133,
8145 (1990); {\it ibid.} {\bf 43}, 8337 (1991).

\bibitem{Moon95}
K. Moon {\it et al},  Phys. Rev. B {\bf 51}, 5138 (1995).


\bibitem{Moore91}
G. Moore and N. Read, Nucl. Phys.
B {\bf 360}, 362 (1991).

\bibitem{bonesteel} N. E. Bonesteel, Phys. Rev. B {\bf 48}, 11484 (1993);
N. E. Bonesteel, I. A. MacDonald, and C. Nayak,
Phys. Rev. Lett. {\bf 77}, 3009 (1996);
M. Greiter, X.-G. Wen, and F. Wilczek, Phys. Rev. Lett.
{\bf 66}, 3205 (1991).

\bibitem{demler}
E. Demler, C. Nayak, and S. Das Sarma, cond-mat/0008137.

\bibitem{rezayi}
E. Rezayi and N. Read, Phys. Rev. Lett. {\bf 72}, 900 (1994); {\it
ibid}. {\bf 73}, 1052 (1994).

\bibitem{girvin}
S. M. Girvin and T. Jach,
Phys. Rev. B {\bf 29}, 5617 (1984).


\bibitem{read94}
N. Read, Semi. Sci. Tech. {\bf 9}, 1859 (1994);
Surf. Sci. {\bf 361}, 7 (1996).

\bibitem{shankar} R. Shankar and G. Murthy,
Phys. Rev. Lett. {\bf 79}, 4437 (1997);
D. H. Lee, Phys. Rev. Lett. {\bf 80}, 4745 (1998);
V. Pasquier and F. D. M. Haldane,
Nucl. Phys. {\bf B 516}, 719 (1998).

\bibitem{stern} B. I. Halperin and A. Stern, Phys. Rev. Lett.
{\bf 80}, 5457 (1998); A. Stern {\it et al.},
Phys. Rev B {\bf 59}, 12547 (1999).

\bibitem{read98} N. Read, Phys. Rev. B {\bf 58}, 16262 (1998).

\bibitem{jain}
J. K. Jain, Phys. Rev. Lett. {\bf 63}, 199 (1989);
Adv. Phys. {\bf 41}, 105 (1992).

\bibitem{CS}
A. Lopez and E. Fradkin,
Phys. Rev. B {\bf 44}, 5246 (1991);
V. Kalmeyer and S.-C. Zhang, Phys. Rev. B {\bf 46},
9889 (1992).

\bibitem{HLR}
B. I. Halperin, P. A. Lee, and N. Read,
Phys. Rev. B {\bf 47}, 7312 (1993).


\bibitem{morinari} T. Morinari, Phys. Rev. Lett. {\bf 81}, 3741 (1998).




\bibitem{green}
N. Read and D. Green,
Phys. Rev. B {\bf 61}, 10267 (2000).


\bibitem{duality}  M. Peskin, Ann. Phys. {\bf 113}, 122 (1978);
P.O. Thomas and M. Stone, Nucl. Phys. B{\bf144},
513 (1978);  C. Dasgupta and B.I. Halperin, Phys. Rev. Lett. {\bf 47},
1556 (1981);  X.G. Wen and A. Zee, Int. J. Mod. Phys. B {\bf 4}, 437 (1990);
 M.P.A. Fisher and D.H. Lee, Phys. Rev. B{\bf 39}, 2756 (1989).

\bibitem{Wen93}
X.G. Wen and Y.S. Wu,  Phys. Rev. Lett. {\bf 70}, 1501 (1993).

\bibitem{Wen99}
X.G. Wen, Phys. Rev. Lett. {\bf 84}, 3950 (2000).

\bibitem{Pryadko94}
L. Pryadko and S.C. Zhang, Phys. Rev. Lett. {\bf 73}, 3282 (1994).

\bibitem{Coleman73}
S. Coleman and E. Weinberg,  Phys. Rev. D {\bf 7}, 1888 (1973);
B.~I. Halperin, T.~C. Lubensky, and S.-k. Ma,
Phys. Rev. Lett. {\bf 32}, 292 (1974).


\bibitem{kim} Y. B. Kim and A. J. Millis,
Physica E {\bf 4}, 171 (1999) and cond-mat/9611125;
I. Ussishkin and A. Stern, Phys. Rev. B {\bf 56}, 4013 (1997);
S. Sakhi, Phys. Rev. B {\bf 56}, 4098 (1997).

\bibitem{vignale} G. Vignale and A. H. MacDonald, Phys. Rev. Lett.
{\bf 76}, 2786 (1996);
I. Ussishkin and A. Stern, Phys. Rev. Lett. {\bf 81}, 3932 (1998);
F. Zhou and Y. B. Kim,
Phys. Rev. B {\bf 59}, R7825 (1999).

\bibitem{yang} K. Yang, Phys. Rev. B {\bf 58}, R4246 (1998).

\bibitem{senthil} T. Senthil and M. P. A. Fisher, Phys. Rev. B
{\bf 60}, 4245 (1999).



\end{references}
\end{document}